\documentclass[twocolumn]{aastex631}
\newcommand{\Ka}{\ensuremath{\hbox{K}\alpha~}}

\def\cha{{\it Chandra }}

\shorttitle{Flux variations and torque reversals of Cen X-3}
\shortauthors{Liu J.}

\begin{document}
\title{Flux variations and torque reversals of Cen X-3}
\correspondingauthor{Jiren Liu}
\email{jrliu@swjtu.edu.cn}
\author{Jiren Liu}
\affiliation{School of Physical Science and Technology, Southwest Jiaotong University, Chengdu Sichuan 611756, China}



\begin{abstract}

Cen X-3 is an archetypical X-ray pulsar with strong flux variations and alternating 
torque reversals, both of which 
are similar to those of recently discovered pulsating ultra-luminous X-ray sources. 
We study a low state of Cen X-3 occurred in 2023 lasting for $\sim100$ days with \cha 
and Insight-HXMT observations, supplemented with MAXI and Fermi/GBM data. 
The \cha spectrum during the eclipse of Cen X-3 in the low state is very similar to 
that in the high state, especially, the Fe lines. 
The HXMT spectrum in the low state shows an enhanced Fe lines, so do the 
MAXI data. The spin-up/spin-down trends of Cen X-3 are not affected by the low states.
All these results indicate that the intrinsic emission in the low states is high, 
and the low states are just apparently low and are dominated by reprocessed emission.
We found that the spin-up to spin-down reversals of Cen X-3 take longer time than
the spin-down to spin-up reversals, which provides a definite observation test
of any possible torque-reversal models. We discuss insights of these results
for understanding the pulsating ultra-luminous X-ray sources.
\end{abstract}


\section{Introduction}

Cen X-3 is the first discovered accretion-powered X-ray pulsar with a spin period
of $\sim4.8\,\rm s$ and an eclipsing orbital period of $\sim$2.1\,days 
\citep{Gia71,Sch72}.
Its optical companion, V779 Cen, is an O6-7 II-III type supergiant \citep{Krz74,Ash99}.
Cen X-3 has played a key role in establishing the physical nature of 
Galactic X-ray sources.
While it has been extensively studied by many X-ray telescopes, some properties 
of Cen X-3 are still not well understood.

Cen X-3 has long been regarded as a disk-fed system, as evidenced by 
its variable optical light curve \citep{Tje86} and the presence of
quasi-periodic oscillations \citep[e.g.][]{Tak91}.
However, no correlation was found between its spin frequency derivative and 
luminosity \citep[e.g][]{Tsu96}.
BATSE on the Compton Gamma-ray Observatory revealed alternating spin-up/spin-down 
intervals of Cen X-3 lasting for tens of days \citep{Fin94,Bil97}. 
A prograde/retrograde disk model 
was proposed to explain its torque reversal \citep{Nel97}.
Recently, it was found that the orbital profile of Cen X-3 peaked at different 
orbital phases for spin-up and spin-down episodes, indicating that its torque reversal 
is related to a process on the orbital scale \citep{LL24}.

On the other hand, the fluxes of Cen X-3 show strong variations, with alternating 
high and low states on time scale of months \citep{Sch76}. The low states could be 
due to absorption or low accretion rate.
Its orbital modulation was found to be intensity-dependent and was suggested to be  
due to varying obscuration of a precessing disk \citep{RP08,Dev10,Bal24}.

In recent years, a few 
ultra-luminous X-ray sources (ULX) were found to show coherent pulsations, unambiguously 
revealed their accretors are magnetized neutron stars \citep[e.g.][]{Bac14}. 
Some properties of 
the pulsating ULXs are very similar to Cen X-3, such as a short orbital period, 
alternating spin-up/spin-down intervals \citep[e.g.][]{Fur23,Liu24}, 
and strong flux variations \citep[e.g.][]{Gur21}. Thus, a better 
understanding of these properties of Cen X-3 would be insightful to understand 
the pulsating ULXs in external galaxies, which are generally much fainter and have less data
available than Cen X-3.

In this paper, we study a low state of Cen X-3 lasting for about 100 days occurred in 2023 with 
\cha and Insight-HXMT observations. The \cha observation was made during an eclipse period
while the HXMT data covered both the high and low states out of eclipse.
We then present an averaged study of different flux level of Cen X-3 with 
MAXI data and discuss the spin-up/spin-down reversal behavior of Cen X-3 using Fermi/GBM 
data. The quoted errors are for 68\% confidence level throughout the paper.

\begin{figure}
	\includegraphics[width=3.5in]{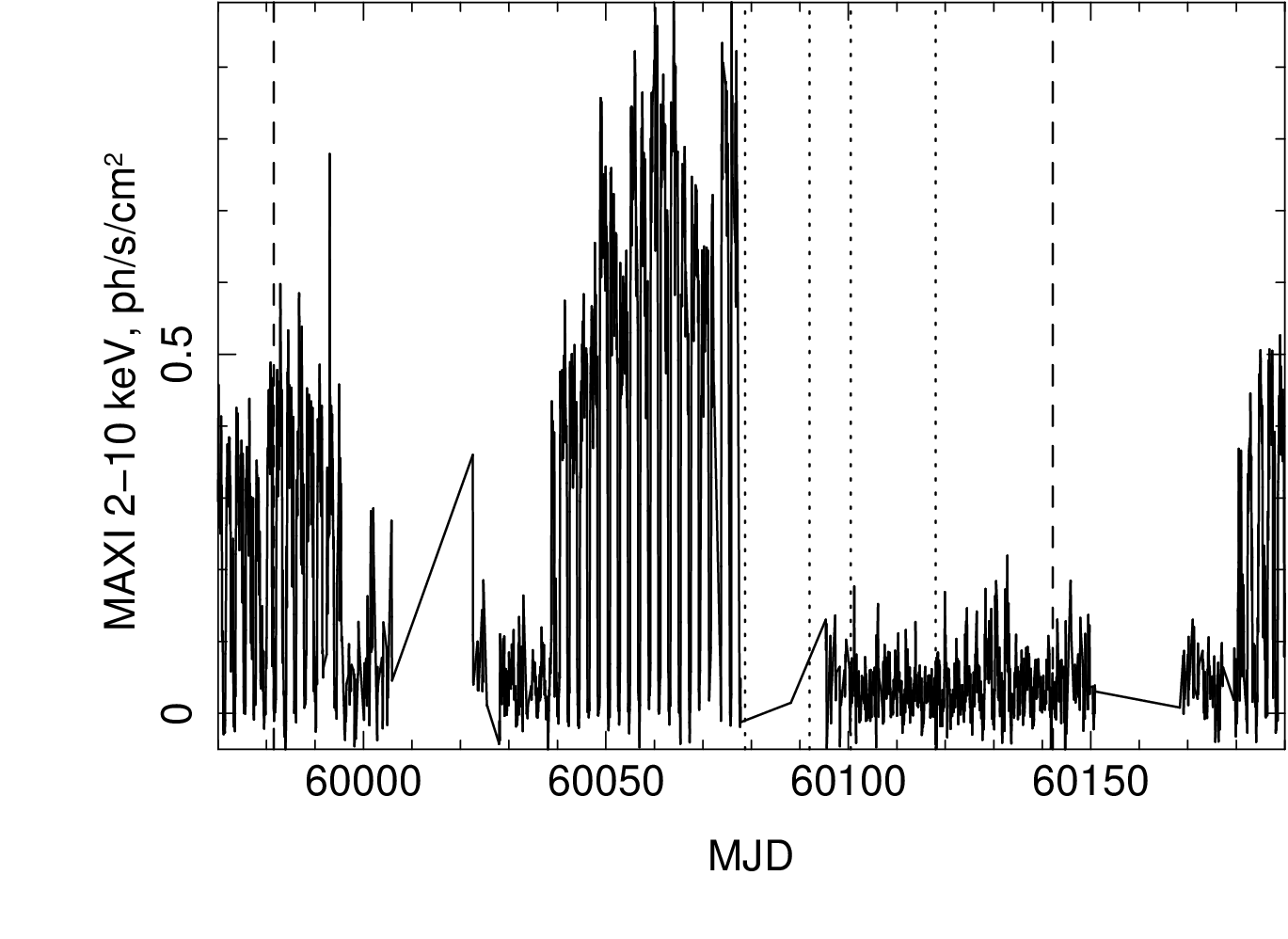}
	\includegraphics[width=3.5in]{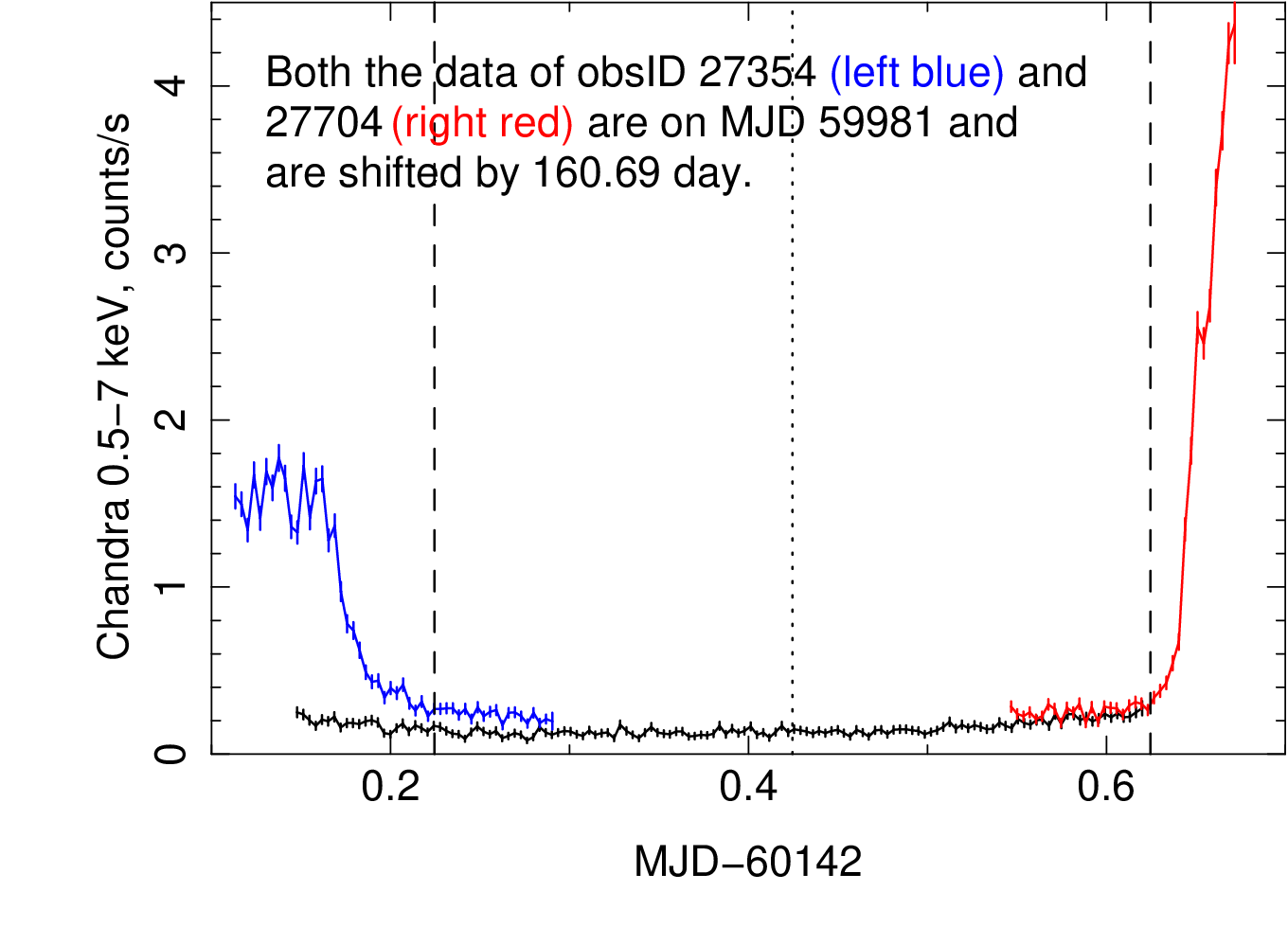}
	\includegraphics[width=3.5in]{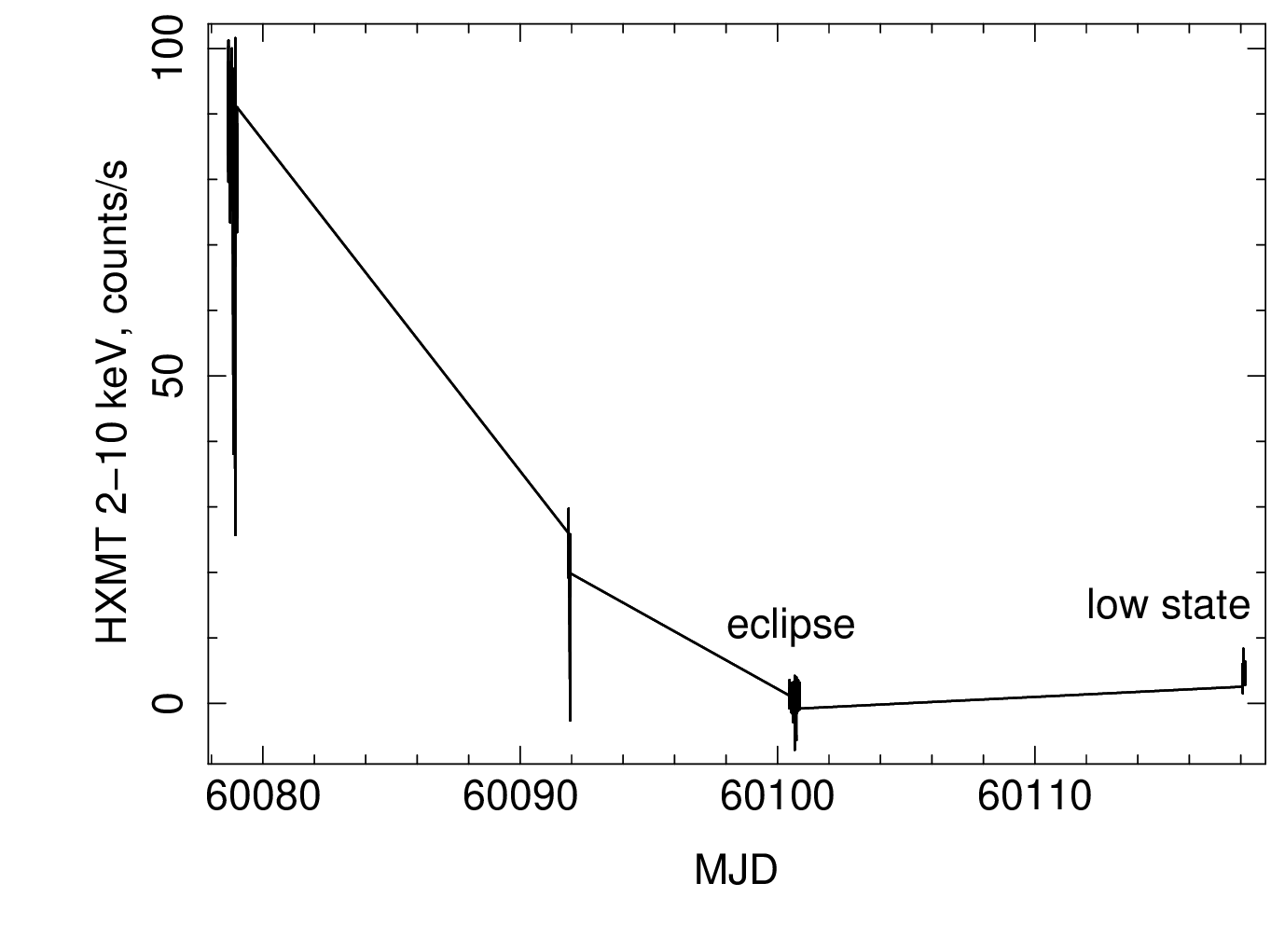}
\caption{Top: light curve of Cen X-3 monitored by MAXI, the vertical dashed lines indicate the 
times of \cha observations and the vertical dotted lines indicates the times of HXMT
observations.
Middle: light curve of three \cha observations, the vertical dashed lines indicate
the time of the eclipse and the dotted line indicates the mid-eclipse.
Bottom: light curve of four HXMT observations.
}
\end{figure}

\begin{figure}
	\hspace{-0.2in}
	\includegraphics[width=3.5in]{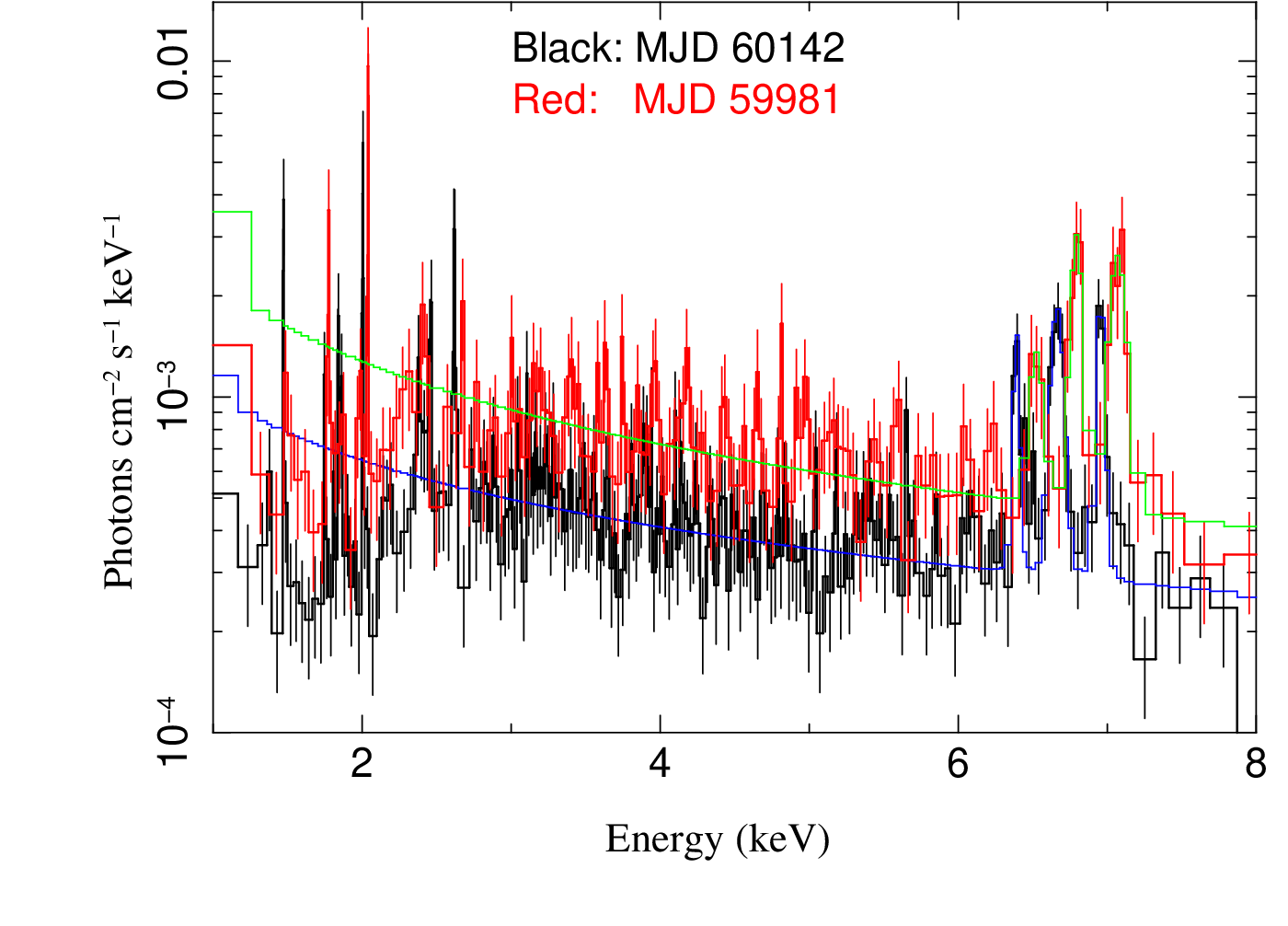}
	\caption{\cha spectrum of Cen X-3 during the eclipse period in a low state (MJD 60142, black) 
and in a high state (MJD 59981, red). For clarity, the red spectrum is right 
shifted a little bit.
}
\end{figure}

\section{Observational data}

Cen X-3 was observed with the High Energy Transmission Grating Spectrometers (HETGs) 
on-board {\it Chandra} on Jul. 17, 2023 (MJD 60142, obsID 26007, PI Canizares),
when it was in a very low state.
This low state lasted for about 100 days, as shown by the MAXI 2-10 keV flux in Figure 1.
For the purpose of comparison, we also analyze two {\it Chandra} observations 
on Feb. 6, 2023 (MJD 59981, obsID 27704 and 27354, PI Canizares) as listed in Table 1. 
The {\it Chandra} data are reduced with CIAO 4.16 following the standard procedure.
We apply a time filter to separate the data into different eclipsing phases.

Insight-HXMT also observed Cen X-3 a few times during the low state around MJD 60110. 
Only one observation (MJD 60118, ObsID P0505126016, PI Liu, Qi) is out of 
eclipse period and is studied here. For comparison we also analyze two high-state 
HXMT observations around MJD 60078 and 60092 and one eclipse observation on MJD 60100.
The observation logs of these observations are also listed in Table 1.

MAXI scans the X-ray sky in the soft X-ray band (2--20\,keV) when it orbits the earth
every time and provides daily fluxes of Cen X-3. Fermi/GBM monitors the pulse frequency 
and pulsed flux of dozens of accreting pulsars in the GBM Accreting Pulsar 
Program \citep{Mal20} and provides a measurement of the spin of 
Cen X-3 for one binary orbital period. 

\begin{table}
\scriptsize
   \begin{center}
\caption{Observation log of \cha and HXMT data}
\begin{tabular}{ccccc}
 \hline
	 Obs.ID & Date & MJD & T$_{eff}$ (ks) & note\\
   \hline
        26007 & 2023-07-17 & 60142 & 41 & eclipse in low state\\
        27354 & 2023-02-06 & 59981 & 15 & partial eclipse in high state\\
        27704 & 2023-02-06 & 59981 & 10 & partial eclipse in high state\\
        0505126010$^a$ &  2023-05-24 & 60078 & 1.4 & high state \\
        0505126012$^a$ &  2023-05-27 & 60092 & 0.7 & middle state\\
        0505126013$^a$ &  2023-06-05 & 60100 & 4.5 & eclipse in low state \\
        0505126016$^a$ &  2023-06-22 & 60118 & 0.7 & low state \\
 \hline
\end{tabular}
\begin{description}
  \begin{footnotesize}
  $^a$For the four HXMT observations, the effective exposure time is for low energy detectors.
  \end{footnotesize}
   \end{description}

\end{center}
\end{table}

\begin{table}
\scriptsize
   \begin{center}
\caption{Fitting results of \cha data}
\begin{tabular}{ccc}
 \hline
	parameter$^a$  & MJD 59981 &  MJD 60142  \\
   \hline
	Norm(PL) & $0.0022\pm0.0008$ & $0.0010\pm0.0004$ \\ 
	$\Gamma$(PL) &  $0.82\pm0.23$ & $0.66\pm0.19$ \\
	F(Fe\,I) (\#/s/cm$^2$) & $0.00009\pm0.00005$ & $0.00006\pm0.00002$ \\
	$E_c$(Fe\,I) (keV) & $6.41\pm0.03$ & $6.39\pm0.02$ \\
	$\sigma$(Fe\,I) (eV) & $38\pm30$ & $12^{+20}_{-12}$ \\
	EW(Fe\,I) (eV)& 190 & 210 \\
	F(Fe\,XXV) (\#/s/cm$^2$) & $0.00023\pm0.00007$ & $0.00016\pm0.00004$\\
	$E_c$(Fe\,XXV) (keV) & $6.68\pm0.01$ & $6.66\pm0.01$ \\
	$\sigma$(Fe\,XXV) (eV) & $30\pm14$ & $35\pm10$ \\
	EW(Fe\,XXV) (eV)&  480 & 540 \\
	F(Fe\,XXVI) (\#/s/cm$^2$) & $0.00028\pm0.00008$ & $0.00011\pm0.00003$\\
	$E_c$(Fe\,XXVI) (keV) & $6.95\pm0.02$ & $6.95\pm0.01$ \\
	$\sigma$(Fe\,XXVI) (eV) & $46\pm17$ & $15\pm11$ \\
	EW(Fe\,XXVI) (eV)& 620 & 380 \\
 \hline
\end{tabular}
\begin{description}
\begin{footnotesize}
$^a$Norm is the normalization of the powerlaw model in units 
of $\rm photon\,keV^{-1}\,cm^{-2}\,s^{-1}$ at 1\,keV, F is the line flux, $E_c$ is the 
energy of line centroid, $\sigma$ is the line width, and EW is the equivalent width.
  \end{footnotesize}
   \end{description}
\end{center}
\end{table}

\begin{figure}
	\hspace{-0.2in}
	\includegraphics[width=3.5in]{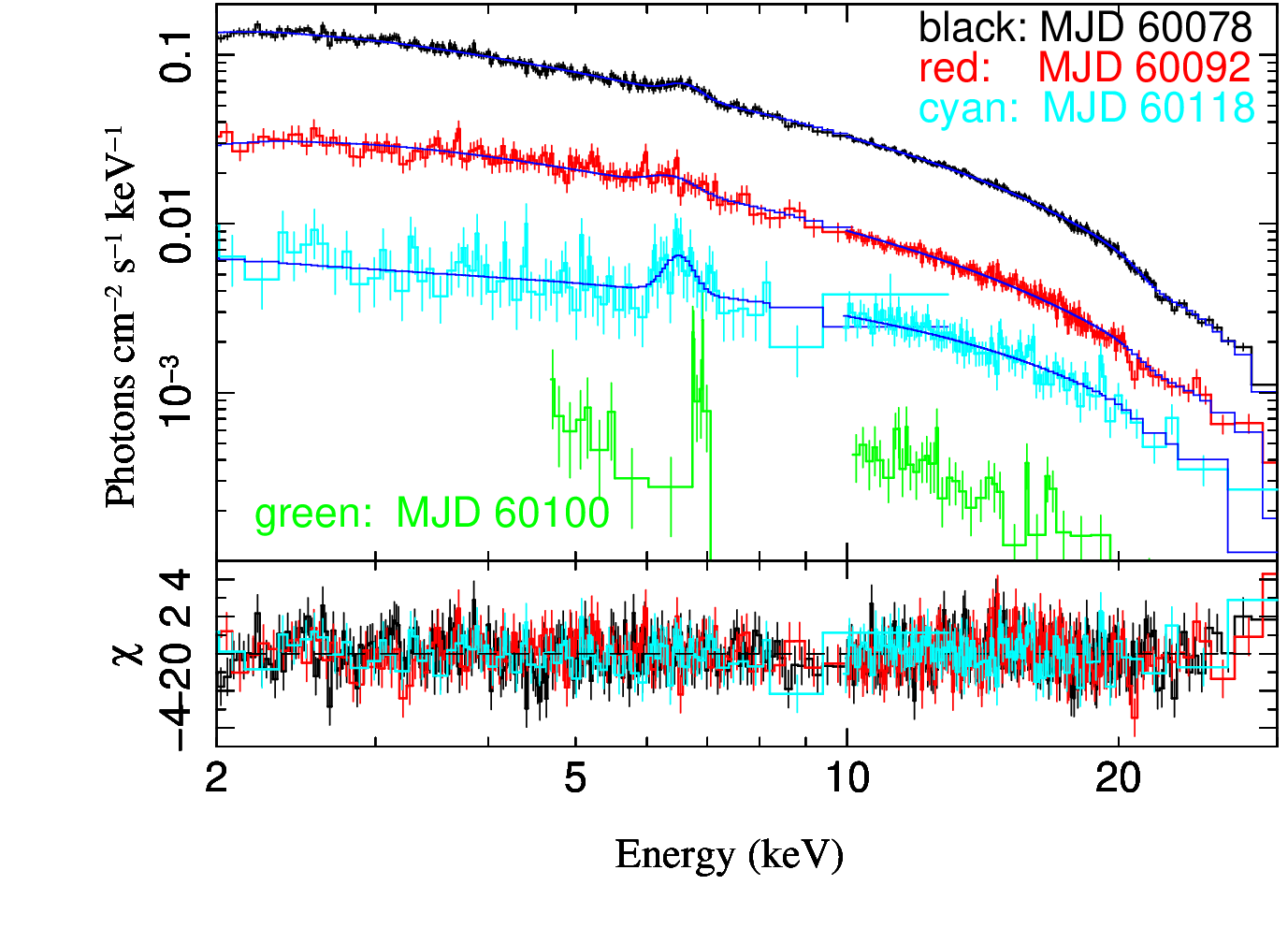}
	\caption{HXMT spectra of Cen X-3 in different states. Black (MJD 60078) is 
in a high state, red (MJD 60092) in a middle state, cyan (MJD 60118) in a low state, 
and green (MJD 60110) for an eclipse period.
}
\end{figure}

\section{Flux variations of Cen X-3}

\subsection{Chandra results}

The lightcurves of all three \cha observations are presented in the middle panel of Figure 1. 
They are extracted from events of both HEG and MEG and both -1 and +1 orders within 0.5-7 keV, 
as the conditions applied in TGCAT 
scripts\footnote{https://tgcat.mit.edu/} \citep{tgcat}. The vertical dotted line indicates 
the predicted mid-eclipse and the two dashed lines indicate the edge of eclipse,
adopting the ephemeris obtained by \citet{Kla23}:
$T_{ecl}=40958.350335$ (MJD),
$P_{\rm orb}=2.087139842$ day,
$\dot{P}_{\rm orb}=-1.03788\times10^{-8}$ day\,day$^{-1}$.
As can be seen, the two observations around MJD 
59981 cover the ingress, egress, and some parts of eclipse, while the observation on MJD 60142
covers ingress and the whole eclipse.

We extract the spectrum of data on MJD 60142 (obsID 26007) during the ingress
(the first 6 ks) and the eclipse (the remaining) separately. We find that 
they look similar. In Figure 2 we plot
the spectrum of the eclipse period as the black histogram. As can be seen, the 
most prominent feature of the spectrum is the neutral-like Fe \Ka line at 6.4 keV, 
and the highly ionized Fe line at 6.7 and 6.95 keV. The corresponding Si lines 
around 2 keV are also clearly shown. For comparison, the eclipse spectrum in a high state
on MJD 59981 (combining both obsID 27354 and 27704) is plotted as the 
red histogram in Figure 2. 

The two spectra look quite similar, and the continuum level of the high state eclipse
(MJD 59981) is about 2 times that of the low state eclipse (MJD 60142).
To quantify the properties of the observed Fe lines, we fit a powerlaw continuum 
plus three Gaussian lines for both spectra within 3-8 keV.
The fitting results are listed in Table 2. The fluxes of the 6.4 keV 
and 6.7 keV Fe lines 
of the low state (MJD 60142) are about 2/3 those of the high state (MJD 59981), 
and the flux of the 6.95 keV Fe line of the low state is about 40\% that 
of the high state. The summed equivalent width (EW) of all three lines 
is about 1.3 and 1.1 keV for MJD 59981 and 60142, respectively.

\begin{figure}
	\includegraphics[width=3.5in]{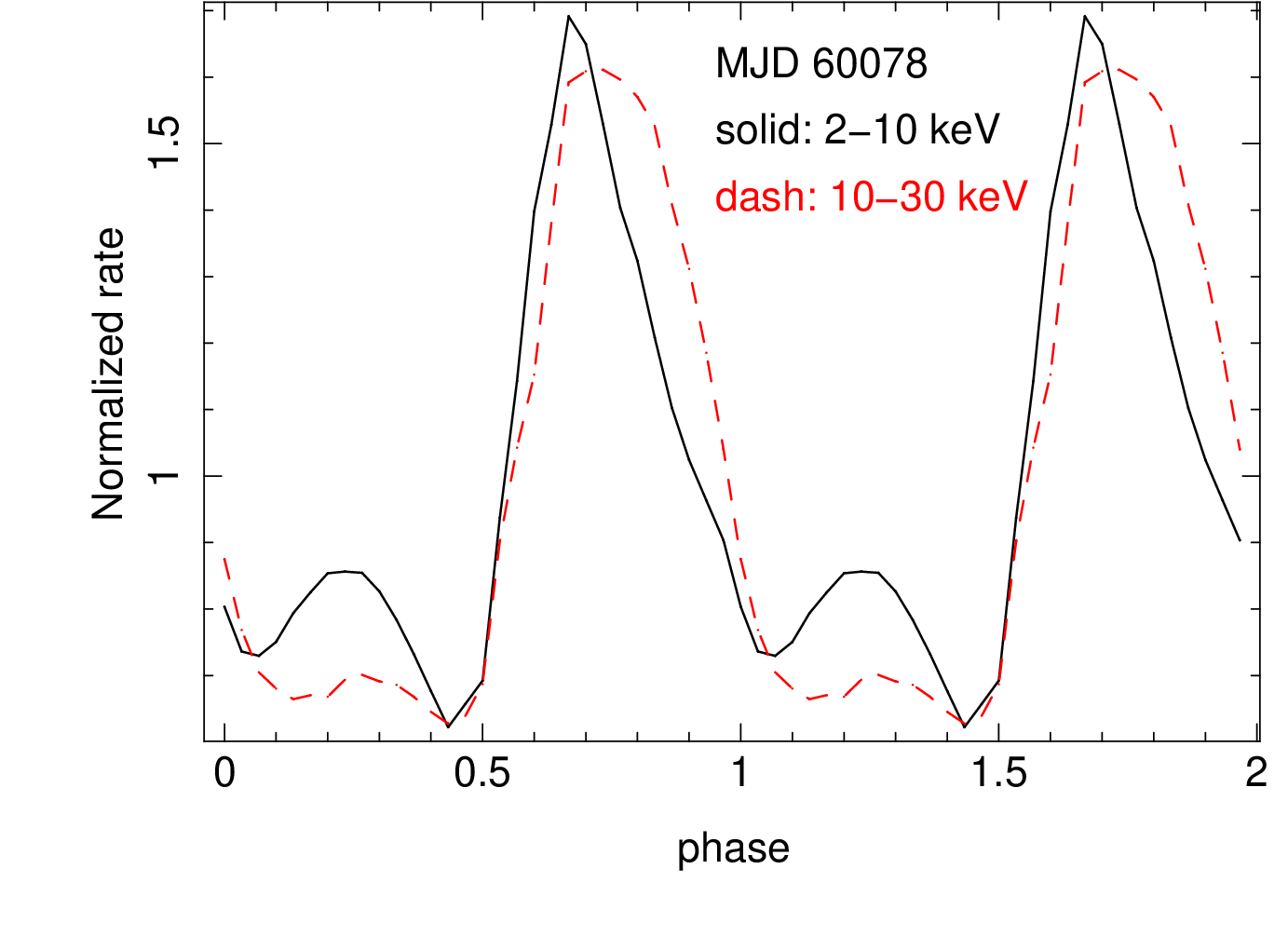}
	\includegraphics[width=3.5in]{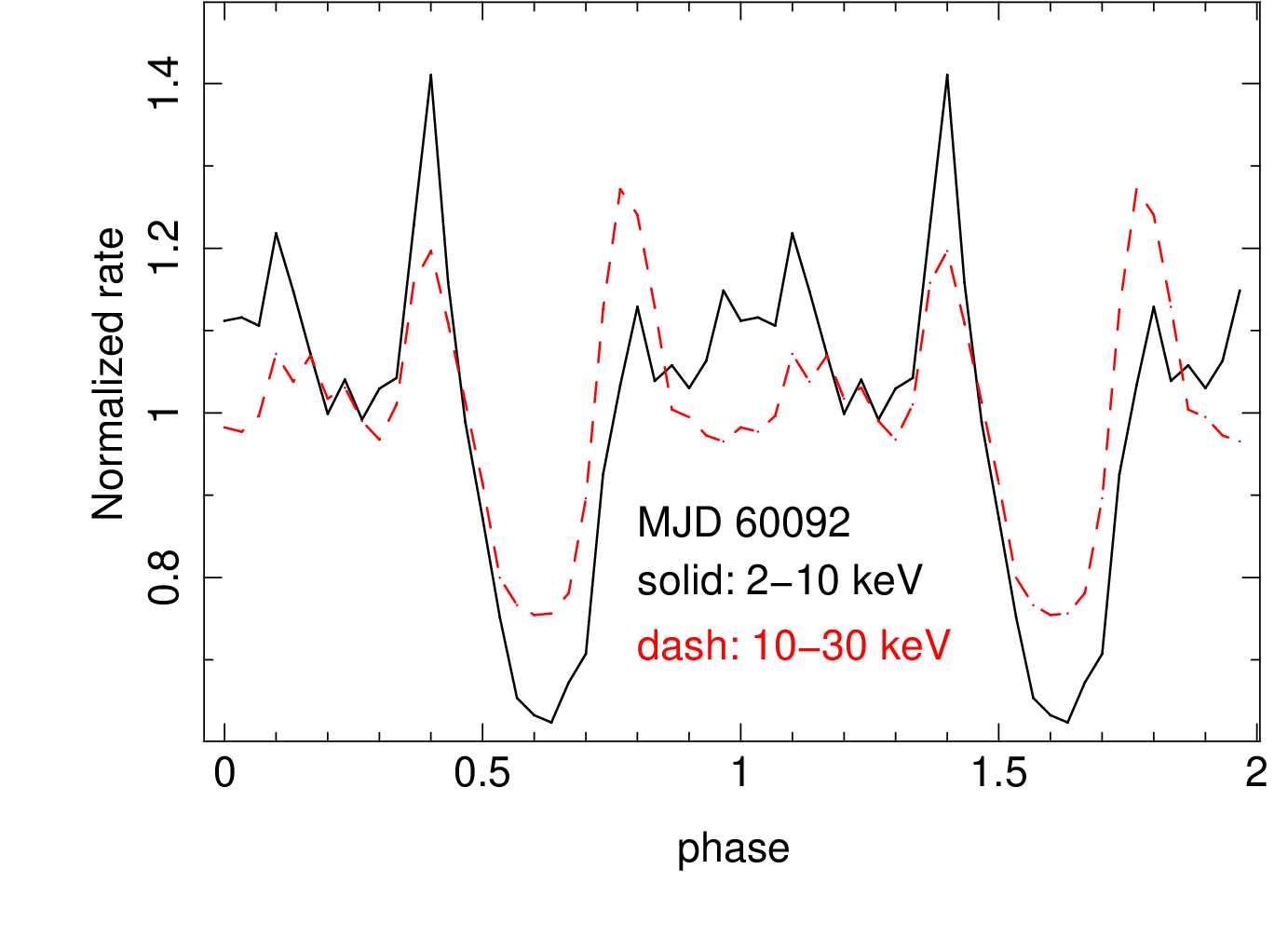}
	\caption{2-10 keV and 10-30 keV pulse profiles of HXMT data in the high state(MJD 60078, 
top) and the middle state (MJD 60092, bottom). No pulsation was detected in 
the low state on MJD 60118.
}
\end{figure}

\begin{table}
\scriptsize
   \begin{center}
\caption{Fitting results of HXMT data}
\begin{tabular}{cccc}
 \hline
	parameter$^a$  & MJD 60078 &  MJD 60092 & MJD 60118  \\
   \hline
	$N_H$ & $1.2\pm0.1$ & $1.4\pm0.5$ & $0^{+0.5}$\\ 
	Norm(PL) & $0.48\pm0.02$ & $0.11\pm0.02$ &$0.007\pm0.002$ \\ 
	$\Gamma$(PL) &  $1.02\pm0.03$ & $0.81\pm0.18$ & $0.19\pm0.17$\\
	$E_{cut}$ &  $15.5\pm0.6$ & $10.7\pm5.1$ & $14.4\pm2.1$\\
	$E_{f}$ &  $5.4\pm0.2$ & $6.8\pm0.6$ & $4.7\pm0.6$\\
	F(Fe\,I) (\#/s/cm$^2$) & $0.006\pm0.001$ & $0.004\pm0.001$ & $0.0014\pm0.0007$\\
	$E_c$(Fe\,I) (keV) & $6.61\pm0.06$ & $6.4^b$ & $6.51\pm0.15$\\
	$\sigma$(Fe\,I) (eV) & $190\pm80$ & $300^b$& $110^{+190}$\\
	EW(Fe\,I) (eV)& 110 & 160 & 400\\
 \hline
\end{tabular}
\begin{description}
\begin{footnotesize}
$^a$$N_H$ is the absorption column density, 
$E_{cut}$ and $E_f$ is the cutoff and folding energy of FDcut model, 
and other parameters are the same as in Table 2.
$^b$The Fe line on MJD 60092 is not well constrained and its centroid energy and width 
are fixed. 
  \end{footnotesize}
   \end{description}
\end{center}
\end{table}

\subsection{HXMT results}
The count rates of the four HXMT observations in 2-10 keV are presented in 
the bottom panel of Figure 1.
The rate of the highest state on MJD 60078 is about 20 times that of the low state
on MJD 60118.

We extract the spectra from all four HXMT observations and they are presented in 
Figure 3. As can be seen, there are Fe line features around 6-7 keV lying on the continua, 
and the lower the spectrum, the more prominent the Fe feature.
The low state spectrum on MJD 60118 is about 20 times less than the high state spectrum 
on MJD 60078 within 2-5 keV, and the factor becomes about 10 within 10-20 keV.
The low state spectrum is about 7 times higher than the eclipse spectrum on MJD 60100.

As did in \citet{LiuQi24} , we fit an absorbed power-law model with Fermi-Dirac
cutoff (FDcut) plus a Gaussian line
to the spectra except for the eclipse spectrum, the photon statistic of which is limited.
The model provides a reasonable fitting to all three spectra, and the fitting results 
are plotted in Figure 3 and listed in Table 3. The fitted photon index of the low state
on MJD 60118 is much flatter than those of the high state on MJD 60078 and 60092.
The Fe line is not well constrained for the spectrum on MJD 60092 and we have fixed 
its centroid energy to 6.4 keV and width to 0.3 keV.
While the continuum of the low state 
on 60118 is about 20 times less than that on 60078, the flux of the Fe line 
is only about 4 times less. The EW of the Fe line in the low state is about 4 times higher than 
that in the high state on 60078.

Besides the spectral features, the pulsation property 
can also tell the nature of the low state.
We converted the photon arrival time to barycentric dynamical time and then corrected 
for binary orbital effect using the orbital parameters obtained by
Fermi/GBM team\footnote{gammaray.nsstc.nasa.gov/gbm/science/pulsars.html}.
We search for a pulsation signal with the epoch folding method in
ISIS \citep{ISIS} with the scripts provided by the Remeis
observatory\footnote{www.sternwarte.uni-erlangen.de/isis}.
We find significant pulsation for the two observations on MJD 60078 and 60092 but 
no pulsation for the low state observation on MJD 60118 and the eclipse observation on MJD 60100.
The pulsation period on MJD 60078 and 60092 is 4.793774\,s and 4.793274\,s, respectively.

The pulsation profiles on MJD 60078 and 60092 within 2-10 keV and 10-30 keV bands are presented
in Figure 4. The low energy profiles look similar as those of high energy profiles, although 
the details show some differences. On the other hand, the high state profiles (MJD 60078)
are totally different from those of the low state profiles (MJD 60092). The high state profiles
are dominated by one peak covering about 0.5 phase, while the low state ones are dominated 
by a plateau covering a phase range of 0.8.
The pulsed fraction of 2-10 keV profiles for MJD 60078 and 60092 is about 0.45 and 0.4, respectively;
while for 10-30 keV profiles, the pulsed fraction is about 0.45 and 0.25, respectively. 

\begin{figure}
	\includegraphics[width=3.5in]{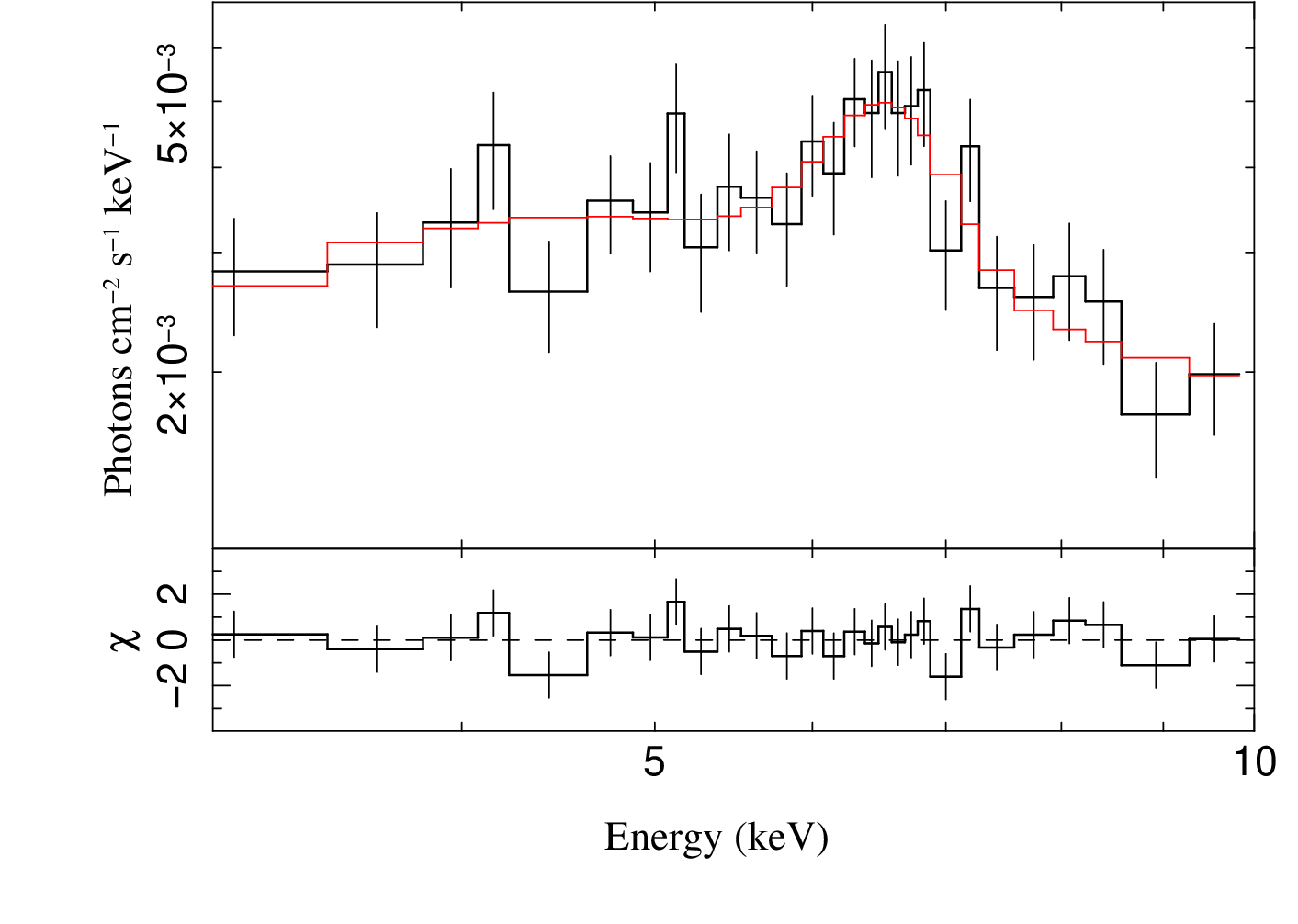}
	\caption{Spectrum of Cen X-3 in the low state between MJD 60095 and 60180 from MAXI data.
The Fe line around 6.5 keV is prominent.
}
\end{figure}
\begin{figure}
	\includegraphics[width=3.5in]{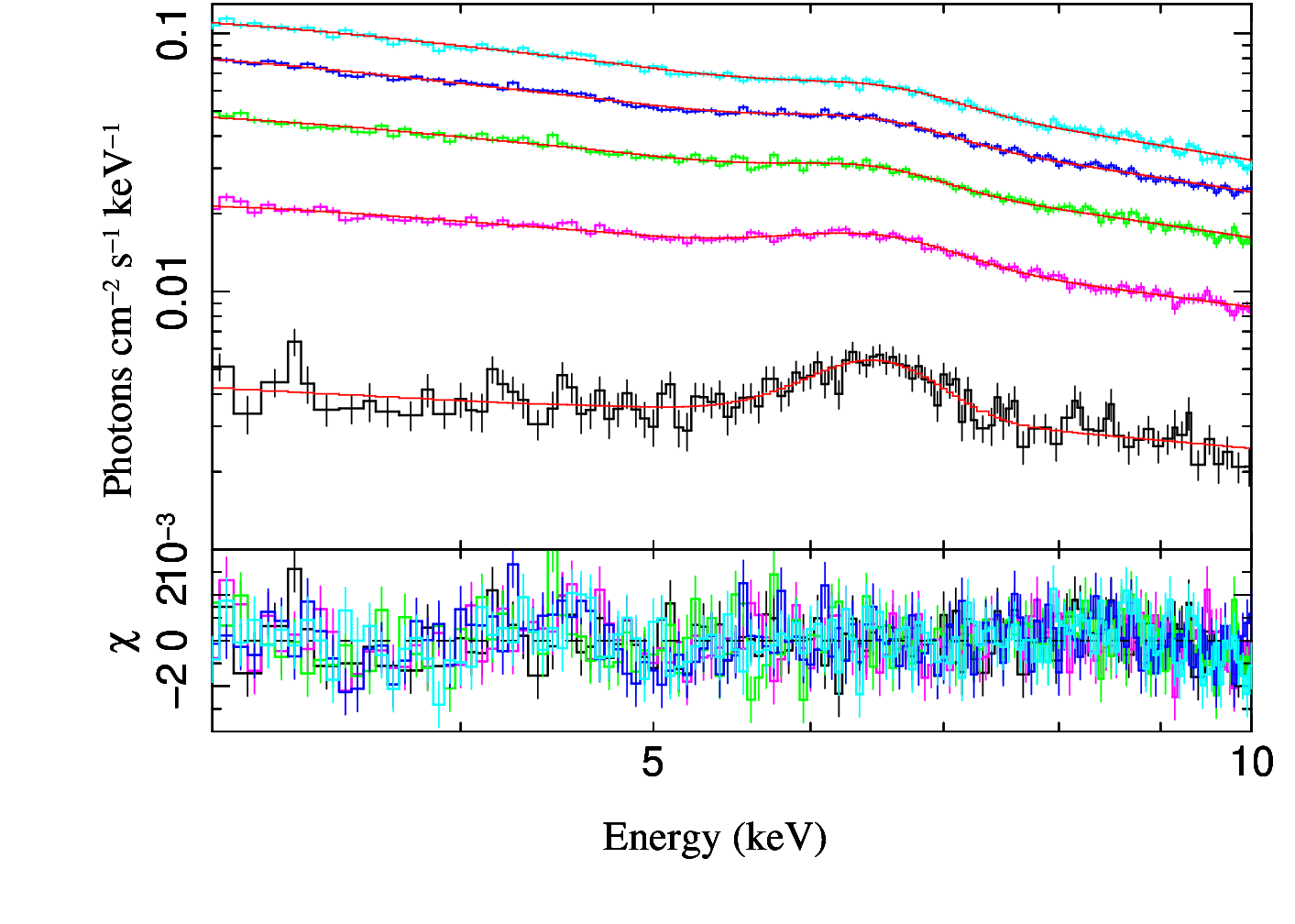}
	\caption{
Spectra of Cen X-3 at five different flux levels from MAXI data.
}
\end{figure}

\begin{figure*}
	\includegraphics[width=6.5in]{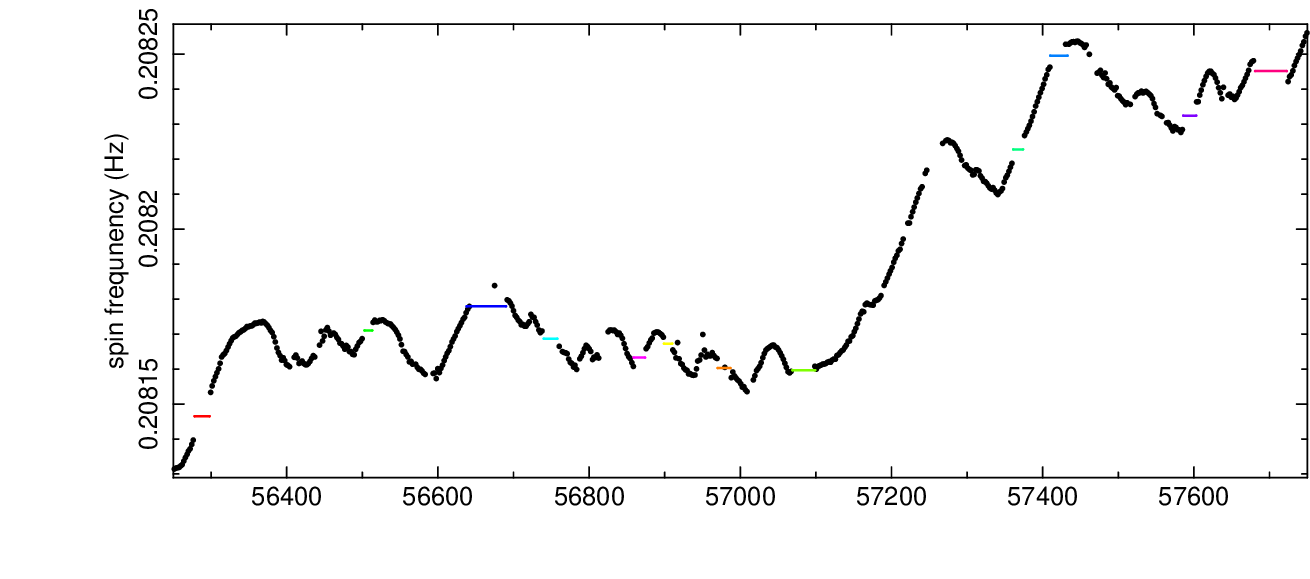}
	\includegraphics[width=6.5in]{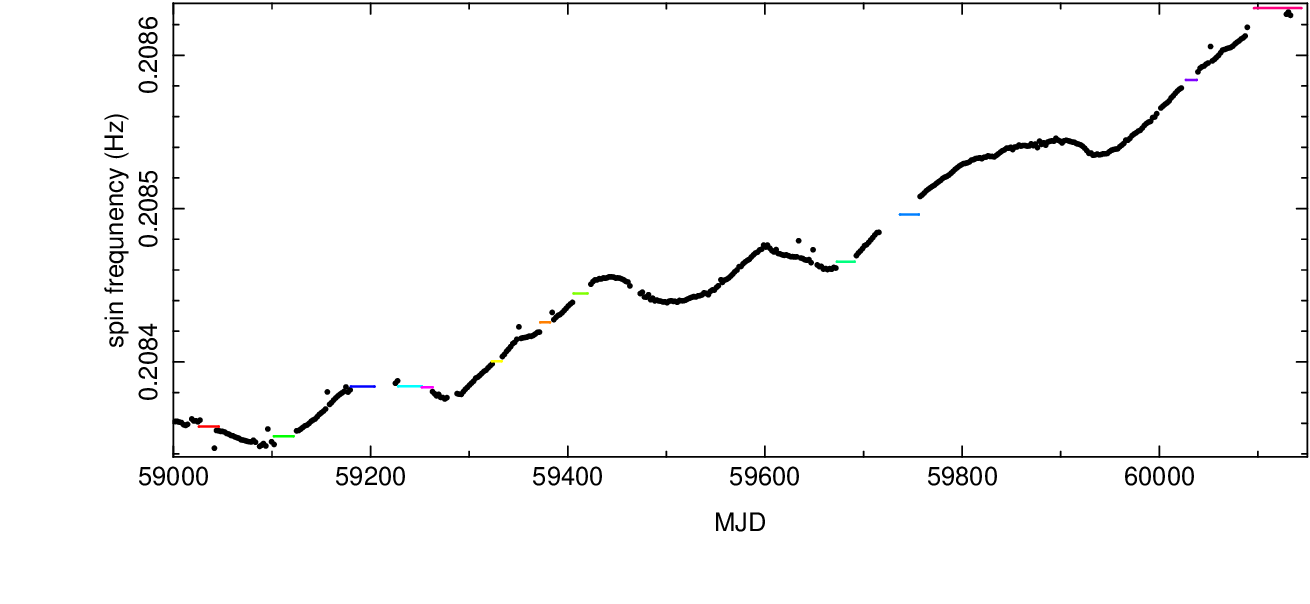}
	\caption{Spin history of Cen X-3 measured by Fermi/GBM in two time periods. 
The horizontal color-bars indicate
the intervals of the identified low states with flux less than 0.1 photons/s/cm$^2$.
}
\end{figure*}

\begin{table*}
\scriptsize
   \begin{center}
\caption{Fitting results of MAXI data}
\begin{tabular}{ccccccc}
 \hline
	parameter$^a$ & MJD 60095-60180	& 0-0.1 & 0.1-0.3 & 0.3-0.5 & 0.5-0.7 & $\ge 0.7$ \\
 \hline
	$N_H$ &$7.0\pm5.3$& $0.95_{-0.95}^{+1.70}$  &$1.79\pm0.30$  &$1.75\pm0.26$ &$1.44\pm0.18$ &$1.78\pm0.24$ \\
	Norm(PL) &$0.03\pm0.01$& $0.010\pm0.003$ &$0.09\pm0.01$  &$0.23\pm0.02$ &$0.39\pm0.02$ &$0.62\pm0.04$ \\ 
	$\Gamma$(PL) &$1.2\pm0.4$& $0.63\pm0.14$   &$1.02\pm0.03$  &$1.15\pm0.03$ &$1.20\pm0.02$&$1.28\pm0.03$\\
	F(Fe) (\#/s/cm$^2$) &$22\pm10$& $25\pm4$ &$46\pm4$ &$50\pm7$ &$79\pm9$ & $101\pm16$\\
	$E_c$(Fe) (keV) &$6.52\pm0.17$& $6.43\pm0.06$     &$6.46\pm0.04$ &$6.41\pm0.06$ &$6.42\pm0.05$&$6.51\pm0.07$ \\
	$\sigma$(Fe) (keV)&$0.01^{+0.54}$ &$0.15\pm0.14$&$0.39\pm0.08$&$0.27\pm0.14$&$0.33\pm0.08$&$0.33\pm0.13$\\
	EW(Fe) (keV)&0.75& 0.83       &0.35 & 0.19&0.20 &0.18 \\
 \hline
\end{tabular}
\begin{description}
  \begin{footnotesize}
  $^a$Parameters are the same as in previous tables.
  \end{footnotesize}
   \end{description}
\end{center}
\end{table*}

\subsection{MAXI results}

As stated above, MAXI monitors the X-ray sky within 2-20 keV daily. This allows 
us to pick up intervals of different flux level of Cen X-3 and do an averaged
spectral study. We first extract the spectrum of the low state between MJD 60095 and 60180
with the on-demand service provided by MAXI team.
The spectrum is presented in Figure 5. The Fe feature around 6.5 keV is very prominent.
As the main spectral feature we focus is the Fe line, 
we fit an absorbed powerlaw plus a Gaussian line to 
the MAXI spectrum within 3-10 keV. The fitted model is plotted as the red line 
in Figure 5 and listed in Table 4. The EW of the Fe line is 
0.75 keV.

We then select continuous intervals longer than 10 days over the whole monitoring 
period of MAXI and divide them into 
five mean flux levels of 0-0.1, 0.1-0.3, 0.3-0.5, 0.5-0.7, and larger than 0.7,
in units of photons/s/cm$^2$ in 2-20 keV band. 
We extract the spectra of different flux levels also with the on-demand service 
provided by MAXI team. The resulting spectra are presented in Figure 6. An emission 
bump around 6.5 keV is also prominent, especially in those of low fluxes.
We also fit these spectra within 3-10 keV. The fitted results are plotted as the red lines 
in Figure 6 and listed in Table 4. 
For the high states, the EW is around 0.2-0.3 keV, but 
for the lowest state, it is about 0.83 keV, similar as the low state between 60095 
and 60180.

\begin{figure}
	\includegraphics[width=3.5in]{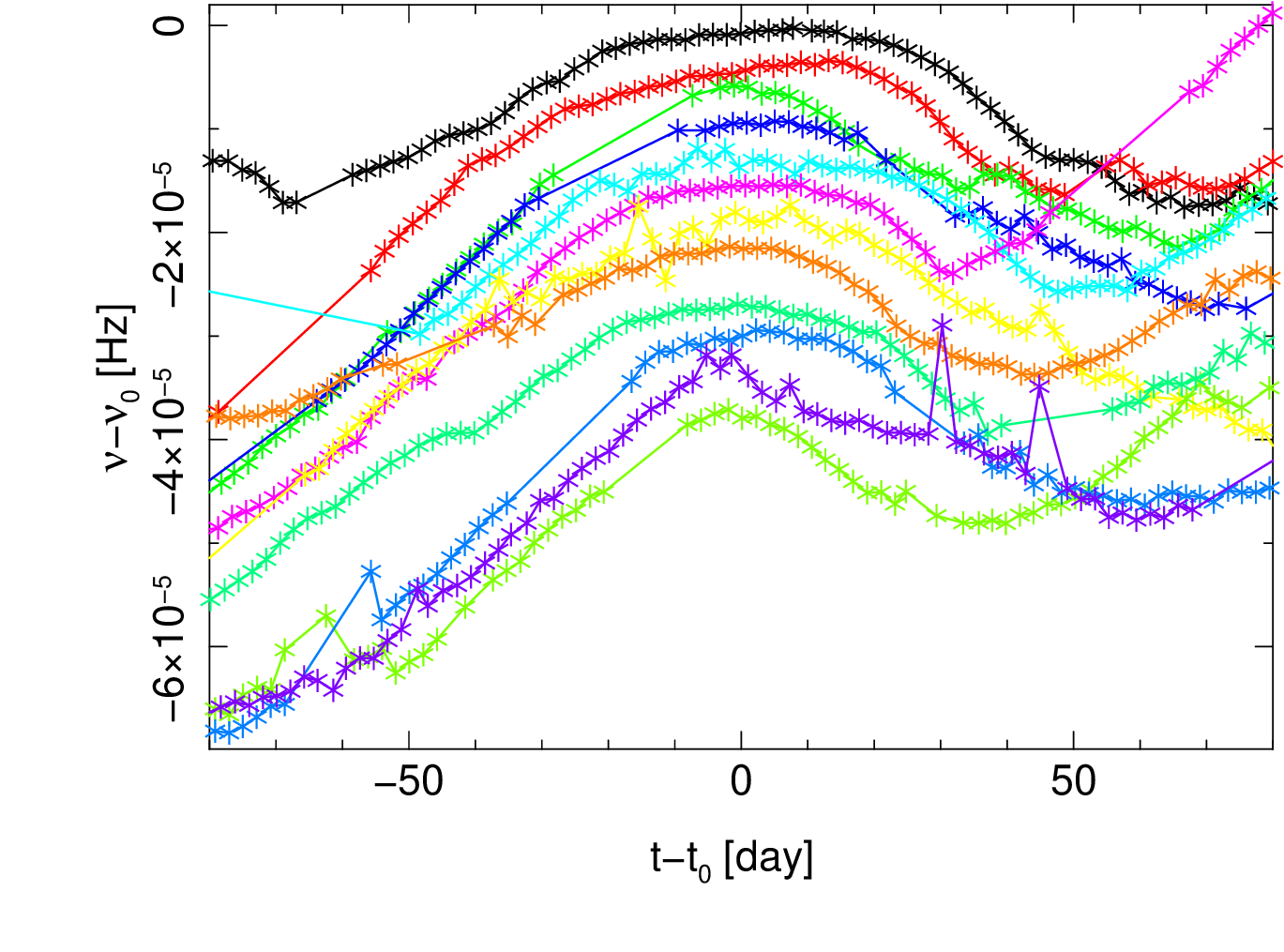}
	\includegraphics[width=3.5in]{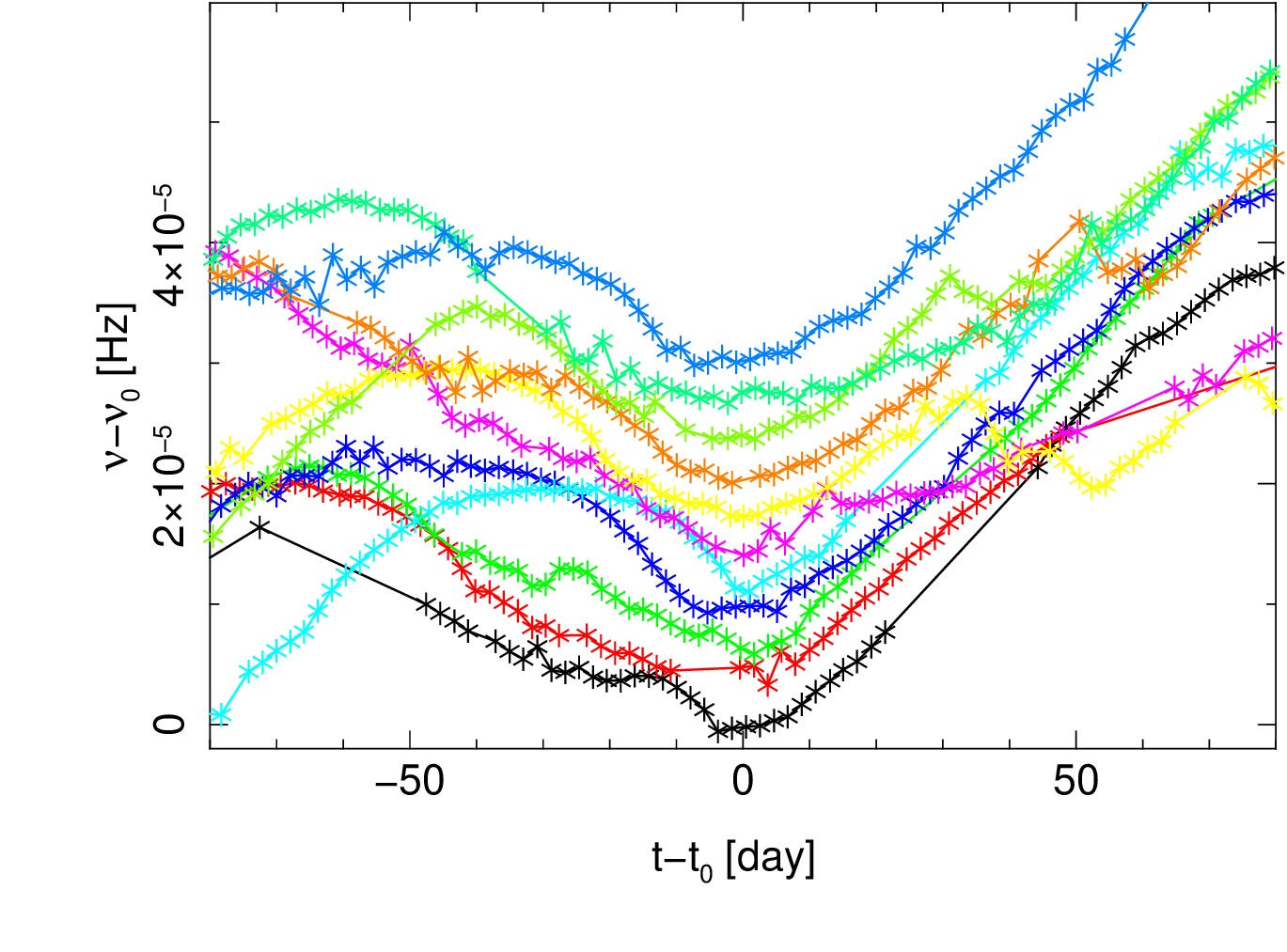}
	\caption{Spin-up to spin-down (top) and spin-down to spin-up reversals of Cen X-3 
measured by Fermi/GBM. The transitions of spin-up to spin-down generally take
longer time than those of spin-down to spin-up.
}
\end{figure}

\subsection{Spin frequency behavior during the low states}

It is also insightful to check the behavior of spin frequency of Cen X-3 during the low
states. 
We totally identified 33 intervals with a mean flux less than 0.1 photons/s/cm$^2$.
Their mean flux is about 0.06 photons/s/cm$^2$. 
Most of the identified intervals have a time length less than 25 days, 
with a few cases lasting for 40-50 days. 
In Figure 7, we plot the Fermi/GBM measured spin frequencies in two periods which 
include about 2/3 of the identified low flux intervals from MAXI.

The spin frequencies of Cen X-3 are erratic. They could fluctuate around some value 
in a few years and could also show an increasing trend with 
alternating spin-up/spin-down periods. 
The gaps of the spin history generally correspond to the relatively low fluxes of Cen X-3. 
The identified low states are indicated as colored horizontal bars in Figure 7.
As can be seen, the low states could happen in spin-up, spin-down, or spin relatively 
stable times, without a preferred spin behavior. For gaps of low fluxes in spin-up/spin-down
episodes, the spin-up/spin-down trend is continuing, not affected by the low fluxes.

\section{Spin reversal behavior of Cen X-3}

In a previous work, we analyzed the orbital profile of Cen X-3 in different 
spin-up/spin-down episodes \citep{LL24}.
Here we present the detailed spin reversal behavior of Cen X-3, that is, 
how spin-up episodes change to spin-down and vice versa.
We identify 12 spin-up to spin-down reversals and plot them together by shifting 
the time and frequency in the top panel of Figure 8. The 11
spin-down to spin-up reversals are presented in the bottom panel.

As can be seen, except for the bottom two cases, most of the spin-up to spin-down reversals 
show a relatively flat transition, lasting for about 30 to 40 days. On the other hand, 
most of the spin-down to spin-up reversals show a rapid transition, with a 
transition period generally less than 15 days.

\section{Discussion and conclusion}

We studied a low state of Cen X-3 lasting for about 100 days around MJD 60095-60180.
The \cha spectrum during the eclipse period in the low state is similar to that 
in the high state, and especially, the Fe lines are similar.
The neutral and highly-ionized Fe lines are produced when neutral and highly-ionized
Fe material are illuminated by X-ray photons higher than their binding energies. 
The similar Fe lines show that the intrinsic emission in the low state is not low,
and the intensity of the low state is just apparently low.

The HXMT observation out of eclipse in the same low state shows a higher 
EW of the Fe line than those in the high and middle states. 
Moreover, in the low state, the HXMT data shows no pulsation signal.
This indicates that the emission in the low state is dominated by a reprocessed 
component, and the intrinsic pulsed signal from Cen X-3 is smeared out.

The averaged MAXI spectrum within MJD 60095--60180 also shows a prominent 
Fe line with an EW of 0.75 keV. Moreover, the MAXI spectrum over all the low states
as observed by MAXI shows a prominent Fe line (with an EW of 0.83 keV), too.
These results show that, in general, the emission in the low states of Cen X-3 is dominated 
by reprocessed emission with a strong Fe line, and the intrinsic emission 
in the low states is not low.
Such a conclusion is also confirmed by the spin frequency behavior of Cen X-3
during the low states, where the increasing/decreasing trends of spin frequency 
are not affected.

The HXMT spectrum in the middle state of Cen X-3 shows no apparent 
absorption at low energies, indicating that in the middle state the intrinsic 
continuum emission is dominating over the reprocessed continuum.
The middle state could be explained as partial coverage of the emission of the 
neutron star, as the partial covering absorption model generally adopted in previous 
work of Cen X-3 \citep[e.g.][]{Nai11}. 
The low-state HXMT spectrum also shows no strong absorption at low 
energies. This indicates that the scattered/reflected soft photons were not heavily 
absorbed. That is, some dense matter obscured the direct emission of the neutron star, 
but not the reprocessed emission. Such a situation would be possible 
if the obscuring matter is asymmetrical, like a warped disk structure, and
the backside/side reflected photons would not be obscured. 

One possibility is the optical companion, which subtends a large solid 
angle ($\sim4\pi/10$) to the pulsar \citep{Moc24}.
A substantial fraction of the X-ray emission could be reflected from the optical star,
and the reflected emission would not be obscured by a smaller structure, such as a 
disk-structure around the pulsar. If this is the case, one would expect a flux peak
when the pulsar lies in the line between the optical star and the earth (orbital phase 0.5), 
The orbital profile
of the low state of Cen X-3 showed a smooth decline after orbital phase 0.5 and 
some plateau before phase 0.5 \citep{RP08}. Considering the changes of the X-ray 
illumination of the 
optical star one can not have a definite answer currently. Nevertheless, it is interesting 
to note that recent XRISM observation of Cen X-3 did show a Fe \Ka profile
peaked around orbital phase 0.5 and the radial velocity profile of the Fe \Ka line
was consistent with an origin of the optical star \citep{Moc24}.

\citet{RP08} has proposed that the low states of Cen X-3 could be due to obscuration 
by a precessing accretion disk. There were some evidence of aperiodic 
timescale around 125-165 days \citep{Pri83}. Recently, \citet{Tor22}
reported a characteristic timescale around 220 days. Such a time scale is not 
as definite as the super-orbital periods in other X-ray pulsars, such as Her X-1.
Nevertheless, we note that all the intervals of low fluxes 
we identified are shorter than these timescales. The accreting matter structure 
of Cen X-3 may be not a perfect disk, but disk-like or even torus-like. 
In principle, the obscuration by such a precessing structure could explain the aperiodic 
low states of Cen X-3.

\citet{San24} reported a transition from the low state to the high state 
around MJD 60038 with \cha observation. They found a Compton shoulder of the Fe \Ka
line short-ward of 6.4 keV and absorption column around $5\times10^{23}$ cm$^{-2}$.
The summed EW of the Fe lines at 6.4, 6.7 and 6.97 keV and the Fe Compton should is 
about 400 eV for their low state spectra (segment 1-a, their Table B1), similar 
to the HXMT value of the low state on MJD 60118. The summed EW reduced to 240 eV and 
110 eV for their high state (segment 2 and 3, their Table B1). 
They considered the transition as the onset of efficient cooling that allows the matter
to enter the magnetosphere. The results we obtained here prefer 
less obscuration for the transition from the low state to the high state. 
They also detected no pulsation 
for their low state data, consistent with our HXMT timing results.

It is interesting to compare the low states of Cen X-3 with those of other X-ray pulsars.
One well-known example is the two low states of Her X-1 during its 35 days super-orbital 
period. Except the super-orbital period of Her X-1 is much stable than Cen X-3, the spectral 
features of the low states of both sources are quite similar. For example, 
the Fe \Ka line was also enhanced in the low states of Her X-1 \citep{AL15} and 
could reach an EW about 0.6 keV \citep{Ji09}; the Fe \Ka line was found 
to be peaked around orbital phase 0.55 and was explained as due to 
reflection from the optical star \citep{AL15}. A power-law component of Her X-1 in the 
low states could be due to scattering from the hot corona of the optical
star \citep{Sha21}.

As we mentioned in the introduction, the pulsating ULXs also show large flux variations.
The flux variations of M82 X-2 has been proposed as due to propeller effect \citep{Tsy16}.
However, a super-orbital period about 60 days was found in M82 X-2 \citep{Bri19}.
Similar super-orbital periods were also found in NGC 7793 P13 \citep{Hu17, Fur18},
NGC 5907 ULX1 \citep{Wal16}, and M51 ULX7 \citep{Vas20,Bri20}.
Such super-orbital periods could be due to a precessing disk as in normal X-ray binaries,
such as Her X-1. The spectra of M82 X-2 and M51 ULX7 show no apparent absorption
in the low states \citep{Bri19,Bri20} and seem to against the explanation of obscuration. 
This is similar to the low state case of Cen X-3 as shown 
in Figure 3. Therefore, the flux variations of pulsating ULXs could be due to variable 
covering fraction of the emission from the neutron star, and the emission 
of their low states could be due to reprocessed emission.
The current X-ray data have no enough signal to reveal the possible Fe line feature
in these extra-galactic sources, and future missions of 
much higher collecting area will help to resolve the real situation. 

Finally, we found that the spin-up to spin-down reversals of Cen X-3 take longer time than 
the spin-down to spin-up reversals. This provides a definite observation test 
of any possible torque-reversal models, such as prograde/retrograde flows. 
In the scenario of flipped inner disk model due to irradiation-driven 
warping instability \citep{van98}, the observer would be viewing the pulsar 
through the disk 
part of the time if the reversal of the disk direction reaches 180 degree, 
which seems to be consistent with the distribution of the low states as shown in Figure 7.
To compare the observed differences in transition times between different reversals
with the scenario, one 
needs to study the distribution of transition times in simulations of flipping disk, and 
one may also need to study their dependence on different parameters, such as the 
viscosity ratio between different directions and the disk X-ray albedo, as mentioned in 
\citet{van98}.
Currently, it is hard to observe the detailed reversal feature in the pulsating ULXs, 
such as M82 X-2 and NGC 5907 ULX1. Cen X-3 is an ideal local analogy 
to understand the reversal behavior of pulsating ULXs. Detailed modelling is needed
to understand the differences between the spin-up to spin-down and
the reverse reversals of Cen X-3.

\section*{Acknowledgements}
We thank the referee for his/her insightful comments
and Hongyuan Zhang for help on spin-reversal of Cen X-3.
This work used data from Insight-HXMT telescope,
MAXI mission, Fermi/GBM and
employed a list of Chandra datasets, obtained by the Chandra X-ray Observatory, 
contained in~\dataset[DOI: cdc.361]{https://doi.org/10.25574/cdc.361}.
This research has made use of a collection of ISIS functions (ISISscripts) provided 
by ECAP/Remeis observatory and MIT.
We acknowledge the support by National 
Natural Science Foundation of China (12473044).

\bibliographystyle{mn2e}

\end{document}